\documentclass [12pt]{article}
\usepackage{graphicx,psfrag}
\usepackage{axodraw}

\catcode`@=12
\newcommand{\be}{\begin{equation}}
\newcommand{\ee}{\end{equation}}
\newcommand{\bea}{\begin{eqnarray}}
\newcommand{\eea}{\end{eqnarray}}

\def\s12{\sin\theta_{12}}
\def\s23{\sin\theta_{23}}
\def\s13{\sin\theta_{13}}
\def\t12{\theta_{12}}
\def\t23{\theta_{23}}
\def\t13{\theta_{13}}
\def\e{\epsilon}

\title{\centerline{\normalsize  SINP/TNP/06-24 \hfill hep-ph/0609193}
{\LARGE \bf Constraining {\it CP} violation in a softly broken 
 $A_4$ symmetric  Model}}
\author{{\bf Biswajit Adhikary\footnote{biswajit.adhikary@saha.ac.in}
    and Ambar Ghosal\footnote{ambar.ghosal@saha.ac.in}}\\
   Saha Institute of Nuclear Physics,\\ 1/AF Bidhan  
        Nagar, Kolkata 700064, India }
\date{}
\begin{document}
\maketitle
\thispagestyle{empty}
\begin{center}
\underbar{Abstract} \\
\end{center}
To understand the mass spectra of charged lepton and neutrino $A_4$
symmetry has been proposed in addition with the Standard
$SU(2)_L\times U(1)_Y$ model. We break $A_4$ symmetry softly and the
deviation from the tri-bimaximal mixing arises due to Zee mechanism.
In the present work, we express two mixing angles $\theta_{13}$ and
$\theta_{23}$ in terms of a single model parameter and experimental
observables, such as, mixing angle $\theta_{12}$, mass squared differences $\Delta
m^2_{21}$ and $\Delta m^2_{32}$. Using the experimental values of
$\theta_{23}$, $\theta_{12}$, $\Delta m^2_{21}$ and $\Delta m^2_{32}$
we restrict the model parameter and we predict $\theta_{13}$. This
model gives rise to $\theta_{13}\simeq 11^\circ$ if we allow $1\sigma$
deviation of $\theta_{23}$ and $2^\circ$ deviation of $\theta_{12}$ from
their best fit values.  Utilizing all those constraints, we explore
the extent of CP violation parameter $J_{\rm CP}$ in the present model
and found a value of $J_{\rm CP}\approx 2.65\times 10^{-3}$ (for
$1\sigma$ deviation of $\theta_{23}$ and $2^\circ$ deviation of
$\theta_{12}$) consistent with the other neutrino experimental
results. We have studied the mass pattern of neutrino and
neutrinoless double beta decay $(\beta\beta_{0\nu})$ parameter $|(M_\nu)_{ee}|$ in this
model.

\newpage
An interesting way to obtain hierarchical charged lepton mass matrix
along with appropriate texture of neutrino mass matrix which can
accommodate present neutrino experimental results, namely, solar,
atmospheric and CHOOZ is through the introduction of non-abelian
discrete $A_4$ symmetry in a model \cite{a4}.  The interplay of $A_4$
symmetry predicts diagonal and hierarchical charged lepton mass matrix
in addition with the neutrino masses which could be quasi-degenerate
or hierarchical. Altarelli and Feruglio have proposed a version of
$A_4$ symmetric model (AF model) \cite{a4af} to obtain hierarchical
charged lepton mass matrix along with tri-bimaximal neutrino mixing
$(\sin\theta_{12} = 1/\sqrt{3},\,\, \sin\theta_{23} = -1/\sqrt{2},
\,\,\s13 = 0)$ \cite{tribi}.  Although the model gives rise to
$\theta_{13}$ = 0 ($|U_{e3}|$ = 0) which is consistent with the
CHOOZ-Palo Verde experimental upper bound ($\theta_{13}<12^\circ$ at
$3\sigma$), however, the non-zero value of $|U_{e3}|$ opens up a
possibility to explore {\it CP} violation in leptonic sector which is
the main goal of many future short and long baseline experiments.

A prediction for non-zero $U_{e3}$ has been realized in a recently
proposed modified AF model through the inclusion of three gauge
singlet charged scalars due to radiative correction of the off diagonal
elements of the neutrino mass matrix \cite{ba}. The model successfully
predicts solar and atmospheric neutrino mixing and mass-squared
differences along with small but non-zero value of $\theta_{13}$ well
below the present experimental upper bound for a reasonable choice of
model parameters. A relationship between different mixing angles is an
outcome of the model and the predictability of the model is also
testable in future neutrino experiments.

Aim of this paper is to generalize the assumptions made in \cite{ba}.  In
the present work we bring down the value of $\theta_{12}$ to its best
fit value $\theta_{12}=34.0^\circ$ from the tri-bimaximal value of the
above $\theta_{12}=35.26^\circ$ which is at the $1\sigma$ edge of
experimental value. We investigate how much $\theta_{13}$ can be
within $1\sigma$ variation of $\theta_{23}$ and at the best fit value
as well as $2^\circ$ variation about the best fit value of
$\theta_{12}$. Then we  shift our concentration to $CP$ violating
parameter $J_{\rm CP}$ and Dirac phase $\delta_D$ and  figure out
their values. We also see the mass pattern of the neutrinos and also 
see the variation of $\beta\beta_{0\nu}$ experimental parameter
in this model. For our analysis we have used the best fit value of the
mass squared differences, $\Delta m^2_{21}=8.0\times 10^{-5}~eV^2$ and
$\Delta m^2_{32}=2.1\times 10^{-3}~eV^2$.

\par
For completeness, we briefly summarize here the model proposed in
\cite{ba}. The lepton content with their representation and the Higgs
content with their vevs and representation under $SU(2)_L\times
U(1)_Y\times A_4$ symmetry is presented in Table 1, where all fields
except the charged scalar $\chi_i^+$ have been used in the original AF model \cite{a4af}. The Yukawa interaction
in the leptonic sector is given by
\begin{eqnarray}
L_{AF}^l &=& y_e e^c(\phi_T l)h_d/\Lambda + 
          y_\mu \mu^c(\phi_T l)^\prime h_d/\Lambda + 
         y_\tau \tau^c(\phi_T l)^{\prime\prime} h_d/\Lambda
         +\nonumber\\
&& x_a\xi(lh_u lh_u)/{\Lambda}^2 + 
        x_b(\phi_S lh_u lh_u)/{\Lambda}^2
\label{laf1}
\end{eqnarray}
where $x_a$, $x_b$, $y_e$, $y_\mu$, $y_\tau$ are Yukawa couplings 
and $\Lambda$ is the new mass scale. After spontaneous breaking of the 
symmetry of the model, the above Lagrangian gives rise to the following 
mass terms
\begin{eqnarray}
{\cal L}_{\rm{AF}}
 &=& v_d v_T/\Lambda ( y_e e^ce + y_\mu \mu^c\mu + y_\tau\tau^c\tau)\nonumber\\
&& +~x_a v_u^2(u/\Lambda^2)(\nu_e\nu_e + 2 \nu_\mu\nu_\tau)\nonumber\\
&& +~x_b v_u^2 2v_S/3\Lambda^2(\nu_e\nu_e + \nu_\mu\nu_\mu\nonumber\\
&& +~\nu_\tau\nu_\tau - \nu_e\nu_\mu - \nu_\mu\nu_\tau - \nu_\tau\nu_e) + h.c.
\label{laf2}
\end{eqnarray}
The charged lepton and neutrino mass matrices come out as
\begin{eqnarray}
m_l = \pmatrix{m_e&0&0\cr
               0&m_\mu&0\cr
               0&0&m_\tau},
m_\nu^{\rm{AF}} = m_0\pmatrix{a + 2d/3&-d/3&-d/3\cr
                                 -d/3&2d/3&a-d/3\cr
                                 -d/3&a-d/3&2d/3},
\label{mlmnu}
\end{eqnarray}
where 
\begin{eqnarray}
m_e = &&y_e v_d v_T/\Lambda, 
m_\mu = y_\mu v_d v_T/\Lambda,
m_\tau = y_\tau v_d v_T/\Lambda,\nonumber\\
&&a = 2x_a u/\Lambda,
d= 2x_b v_S/\Lambda, 
m_0=v_{\rm u}^2/\Lambda.
\label{lp}
\end{eqnarray}
The charged lepton mass matrix is diagonal so the leptonic mixing 
solely occurs from the neutrino sector and diagonalising $m_\nu$ 
by the way $U^\dagger m_\nu U^\star$ we get the three mass eigenvalues 
as 
\begin{eqnarray}
m_1 = a+d , m_2 = a, m_3 = d-a
\label{mev}
\end{eqnarray}
with the exact tri-bimaximal mixing pattern. Thus, in the AF model,
tri-bimaximal mixing occurs naturally. In \cite{ba}, a modified
version of the above model has been investigated to generate non-zero
$\theta_{13}$ which at the leading order gives tri-bimaximal mixing
form. In order to do that, an $A_4$ triplet $SU(2)_L$ singlet charged
scalar $\chi_i^+$(= $\chi_1^+,\chi_2^+,\chi_3^+$) has been
incorporated and the leptonic part of the Lagrangian becomes
\begin{eqnarray}
L = L_{AF}^l + L_{MAF}^l
\end{eqnarray}
where $L_{MAF}^l$ has two parts as 
\begin{eqnarray}
L_{MAF}^l = {\cal L}_1 + {\cal L}_2
\end{eqnarray}
${\cal L}_1$ is $A_4$ symmetry preserving part and is given by 
\begin{eqnarray}
&&{\cal L}_1 = f~(L~L~\chi_i) \subset (3 \times 3 \times 3)\nonumber\\
&&= f(\nu_\mu \tau \chi^+_1+ \nu_\tau e \chi^+_2 +
\nu_e \mu \chi^+_3
-\nu_\tau \mu \chi^+_1- \nu_e \tau \chi^+_2
- \nu_\mu e \chi^+_3
).
\label{maf1}
\end{eqnarray}
however, ${\cal L}_2$ contains Zee-type term which is explicit soft 
$A_4$ symmetry breaking and is given by 
\begin{eqnarray}
{\cal L}_2 = c_{12}h_u^T i\tau_2 h_d(\chi_1^+ + \chi_2^+ + \chi_3^+).
\end{eqnarray}
The charged lepton mass matrix is still diagonal in the present model and the neutrino mass matrix comes out as
\begin{eqnarray}
m_\nu = \pmatrix{a + 2d/3&-d/3&-d/3 - \epsilon\cr
                       -d/3&2d/3&a-d/3 + \epsilon\cr
                       -d/3 - \epsilon&a-d/3 +\epsilon&2d/3}
\label{mnu1}
\end{eqnarray}
where the $a$ and $d$ parameters defined earlier 
and are obtained due to higher dimensional 
operators in the same way as AF model. The $\epsilon$ term is the 
additional contribution arises at the one-loop level due to well 
known Zee mechanism and is shown in Fig.1. The parameter 
$\epsilon$ is given by 
\begin{eqnarray}
\epsilon = fm_\tau^2 {\frac{c_{12}v_u}{v_d}}F(m_\chi^2, m_{h_d}^2)
\end{eqnarray}
with the definition,
\begin{eqnarray}
F(M_1^2,M_2^2) = {\frac{1}{16\pi^2(M_1^2-M_2^2)}}
\ln{\frac{M_1^2}{M_2^2}}.
\end{eqnarray}
Although the Lagrangian given in Eq.\ref{maf1} can generate corrections 
to all off-diagonal entries of the mass matrix given in Eq.\ (\ref{mnu1}), 
however, dominant terms proportional to $m_\tau^2$ are retained. 
In this model neutrinos are Majorana-type in nature. In general, the 
parameters $a$, $d$, $\epsilon$ are all complex, however, it is 
possible to rotate out one of the phase. For our analysis, we consider 
only the parameter $d$ is complex and parameters $a$,$\epsilon$ are 
real. The neutrino mass matrix in this case comes out as
\begin{figure}[htbp]
\begin{center}\begin{picture}(300,100)(0,45)
\ArrowLine(50,50)(110,50)
\ArrowLine(150,50)(110,50)
\ArrowLine(190,50)(150,50)
\ArrowLine(250,50)(190,50)
\Text(80,55)[b]{$\nu_{e,\mu}$}
\Text(220,55)[b]{$\nu_{\tau}$}
\Text(130,55)[b]{$\tau_L$}
\Text(150,50)[]{$\times$}
\Text(170,55)[b]{$\tau_R$}
\Text(115,80)[b]{$\chi^+_{2,1}$}
\Text(185,80)[b]{$h_d^+$}
\DashLine(150,90)(150,115){2}
\Text(155,117)[b]{$<h_u^0>$}
\DashCArc(150,50)(40,0,180){2}\Vertex(110,50){2}\Vertex(190,50){2}
\end{picture}\end{center}
\caption[]{\label{fen}One loop radiative $\nu_{e,\mu}$ - $\nu_\tau$ mass
due to charged Higgs exchange.}\end{figure}
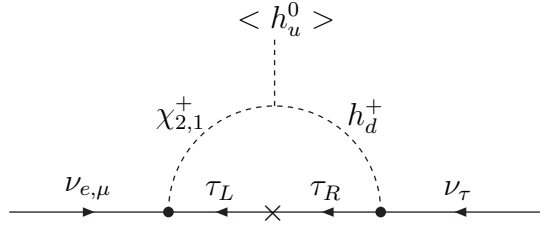
\begin{table}
\begin{center}
\begin{tabular}{|c|c|c|c|}
\hline
{\rm Lepton}& $SU(2)_L$ & $A_4$&\\
\hline
$(\nu_i, l_i)$&2&3&\\
$l_i^c$&1&1&\\
\hline
Scalar&&&{\rm VEV}\\
\hline
$h_u$&2&1&$<h_u^0>$= $v_u$\\
$h_d$&2&1&$<h_d^0>$=$v_d$\\
$\xi$&1&1&$<\xi^0>$ = u\\
$\phi_S$&1&3&$<\phi_S^0>$ = $(v_S,v_S,v_S)$\\
$\phi_T$&1&3&$<\phi_T>$ = $(v_T,0,0)$\\
$\chi_i^+$&1&3&\\
\hline
\end{tabular}
\caption{List of fermion and scalar fields used in this  model.
}
\end{center}
\end{table}
%
\begin{eqnarray}
 M_\nu = \pmatrix{a + 2de^{i\phi}/3&-de^{i\phi}/3&-de^{i\phi}/3-\e\cr
                  -de^{i\phi}/3&2de^{i\phi}/3&a-de^{i\phi}/3+\e\cr
                  -de^{i\phi}/3-\e&a-de^{i\phi}/3+\e&2de^{i\phi}/3}.
\label{mnumat}
\end{eqnarray}
 A relationship between
the parameters $a$ and $d$ has been considered as
\begin{eqnarray}
d= \kappa a \cos\phi
\label{darel}
\end{eqnarray}
where $\kappa$, $d$ and $a$ are the real parameters. The analysis of
Ref.\cite{ba} has been done for a specific value of $\kappa$ and in
the present work, we have generalized the whole analysis. We constrain
the value of $\kappa$ first from the existing bounds of two mixing
angles $\theta_{12}$ and $\theta_{23}$. Then we have utilized the
result to calculate $\theta_{13}$ and the CP
violation parameter $J_{\rm CP}$ \cite{branco1} and explore the extent
at which $J_{\rm CP}$ is allowed in the present model. 
 
With only assumption $\e$ is small, we diagonalize mass matrix Eq.\ (\ref{mnumat}) as
\begin{eqnarray}
U^\dagger M_\nu U^*={\rm diag}\left(de^{i\phi}+a+\e,~a,~de^{i\phi}-a-\e\right)
\label{md}
\end{eqnarray}
upto first order in $\e$ where diagonalizing matrix
\begin{eqnarray}
U&=& \left(\begin{array}{ccc}\sqrt{\frac{2}{3}} & 
\sqrt{\frac{1}{3}} &
0\cr
-\sqrt{\frac{1}{6}}& 
\sqrt{\frac{1}{3}} &-\sqrt{\frac{1}{2}} \cr
     -\sqrt{\frac{1}{6}} &\sqrt{\frac{1}{3}} & \sqrt{\frac{1}{2}} \end{array}\right)+\nonumber\\
 &&\e\left(\begin{array}{ccc}-\frac{2a +
    de^{-i\phi}}{\sqrt{6}d(d+2a\cos\phi)} & 
\frac{2a +
    de^{i\phi}}{\sqrt{3}d(d+2a\cos\phi)} &
\frac{1}{3\sqrt{2}}\left(\frac{1}{a}-
\frac{e^{-i\phi}}{d-2a\cos\phi}\right)\cr
\frac{
  d^2-4a^2+2iad\sin\phi}{2\sqrt{6}ad(d+2a\cos\phi)}& 
\frac{2a^2\cos\phi-d(a+de^{i\phi})
    }{\sqrt{3}(d^3-4a^2d\cos^2\phi)} &-\frac{1}{6\sqrt{2}}\left(\frac{1}{a}+
\frac{2e^{-i\phi}}{d-2a\cos\phi}\right) \cr
     -\frac{4a^2
  +d^2+4ad\cos\phi-2iad\sin\phi}{2\sqrt{6}ad(d+2a\cos\phi)}             &\frac{a(2a\cos\phi+de^{2i\phi})
    }{\sqrt{3}(d^3-4a^2d\cos^2\phi)}& -\frac{1}{6\sqrt{2}}\left(\frac{1}{a}+
\frac{2e^{-i\phi}}{d-2a\cos\phi}\right)\end{array}\right).
\label{U}
\end{eqnarray}
It is to be noted that for vanishing value of $\epsilon$ the matrix
$U$ leads to tri-bimaximal form. Three approximate mass eigenvalues
in Eq.\ (\ref{md}) take the following forms
\begin{eqnarray}
{m_1}^2  &=& \left|de^{i\phi}+a+\e\right|^2\nonumber\\ &\simeq& a^2\left[1 + 2\e^\prime + 2\kappa\cos^2\phi + 
          2\e^\prime\kappa\cos^2\phi +
          \kappa^2\cos^2\phi\right]\nonumber\\
{m_2}^2 &=& a^2\nonumber\\
{m_3}^2 &=& \left|de^{i\phi}-a-\e\right|^2\nonumber\\&\simeq& a^2\left[1+2\e^\prime - 2\kappa\cos^2\phi - 
          2\e^\prime\kappa\cos^2\phi + \kappa^2\cos^2\phi\right]
\label{mev}
\end{eqnarray}
where $\e^\prime = \e/a$ and we use the relation given in 
Eq.\ (\ref{darel}).
The solar and atmospheric neutrino mass squared differences are coming
out as
\begin{eqnarray}
\Delta m^2_\odot &=& \Delta m^2_{21}=m_2^2 - m_1^2 
= a^2\left[-\kappa\cos^2\phi(\kappa+2) - 
2\e^\prime(1+\kappa\cos^2\phi)\right]\nonumber\\
\Delta m^2_{\rm atm}& =&\Delta m^2_{32}= m_3^2 - m_2^2
= a^2\left[-\kappa \cos^2\phi(2-\kappa) + 2\e^\prime
(1-\kappa\cos^2\phi)\right].
\label{massdif}
\end{eqnarray}
From mixing matrix $U$ in Eq.\ (\ref{U}) we obtain the mixing angles:
\begin{eqnarray}
  &&\sin\theta_{12} =\left|U_{12}\right|= \frac{1}{\sqrt{3}} + \frac{\e^\prime(2+\kappa\cos^2\phi)}{\sqrt{3}\kappa\cos^2\phi(\kappa+2)}\nonumber\\
&&\sin\theta_{13} =\left|U_{13}\right|= \left|\frac{\e^\prime}{3\sqrt{2}(\kappa-2)}\right|\left[
\frac{1+\cos^2\phi(\kappa^2+8 - 6\kappa)}{\cos^2\phi}\right]^{1/2}\nonumber\\
&&\tan^2\theta_{23} =\frac{\left|U_{23}\right|^2}{1-\left|U_{23}\right|^2}= 1 +\frac{2\e^\prime\kappa}
{3(\kappa-2)}.
\label{angles}
\end{eqnarray}
We define the ratio $R$ as 
\begin{eqnarray}
R = \frac{\Delta m^2_\odot}{\Delta m^2_{atm}} =
\frac{\kappa\cos^2\phi(\kappa+2)+2\e^\prime(1+\kappa\cos^2\phi)}
{\kappa\cos^2\phi(2-\kappa) - 2\e^\prime(1-\kappa\cos^2\phi)}.
\label{r}
\end{eqnarray}
Using the first relation of Eq.\ (\ref{angles}) we have
\begin{eqnarray}
\e^\prime =\frac{\sqrt{3}\kappa\cos^2\phi(\kappa+2)}{2+\kappa\cos^2\phi}\times\left(\sin\theta_{12} - 1/\sqrt{3}\right).
\label{eprime}
\end{eqnarray}
Using  Eq.\ (\ref{eprime}) in Eq.\ (\ref{r}) we get
\begin{eqnarray}
\cos^2\phi = \frac{-2}{\kappa}\left[\frac{2(1-R)+\kappa(1+R)+\sqrt{3}(\kappa+2)(\sin\theta_{12} - 1/\sqrt{3})(1+R)}
{2(1-R) + \kappa(1+R)+2\sqrt{3}(\kappa+2)(\sin\theta_{12}-1/\sqrt{3})(1-R)
}\right].
\label{cosphi}
\end{eqnarray}
%
\begin{figure}
\begin{center}
\includegraphics[height=8cm,keepaspectratio]{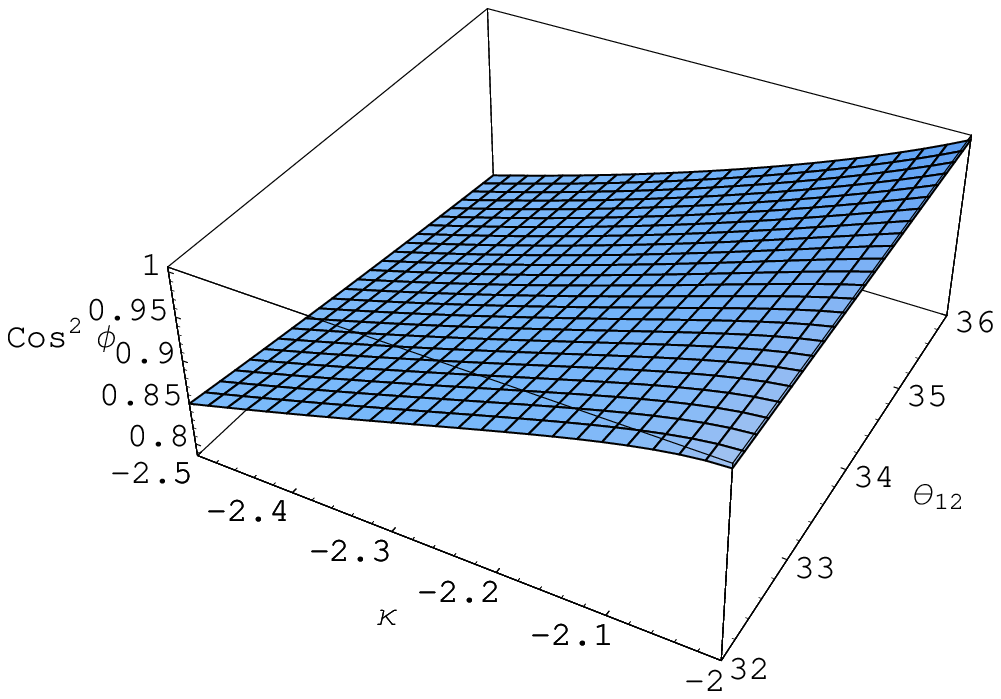}
\caption{\label{cphi} Plot of   $\cos^2\phi$  with respect to
  $\kappa$ and  $\theta_{12}$. We keep
  $\Delta m^2_{32}$ and  $\Delta m^2_{21}$ to their
  best fit values.  }
\end{center}
\end{figure}
%
Again using expression of $\e'$ from Eq.\ (\ref{eprime}) to the
expression of $\Delta m_{32}^2$ in Eq.\ (\ref{massdif}) we get
the dependence of the $a^2$ on $\kappa$ and $\cos^2\phi$:
\begin{eqnarray}
a^2= \frac{\Delta
m_{32}^2(2+\kappa\cos^2\phi)}{\kappa\cos^2\phi[(\kappa-2)(2+\kappa\cos^2\phi)+2\sqrt{3}(\kappa+2)(1-\kappa\cos^2\phi)(\sin\theta_{12}-1/\sqrt{3})]}
\label{a}
\end{eqnarray}
Thus from Eq.\ (\ref{cosphi}) and Eq.\ (\ref{a}), we see that $a^2$
only depends on single model parameter $\kappa$.  From the above
expression of $\cos^2\phi$ we can put bound on $\kappa$ for the given
values of $R$ and $\theta_{12}$. For the best fit value of $\Delta m_{32}^2$
and $\Delta m_{21}^2$ we have the approximate bound:
\begin{eqnarray}
\kappa < -2
\label{kbound1}
\end{eqnarray}
in the course of variation $32^\circ\le\theta_{12}\le 36^\circ$ for
$\cos^2\phi<1$ and it is shown in Fig. \ref{cphi}. It also ensures $\cos^2\phi>0$.  Again
compatibility of the above bound of $\kappa$ with the restriction
$a^2>0$ demands that $\Delta m_{32}^2>0$. This leads to normal
ordering of neutrino masses.

\begin{figure}
\begin{center}
\includegraphics[height=10cm,keepaspectratio]{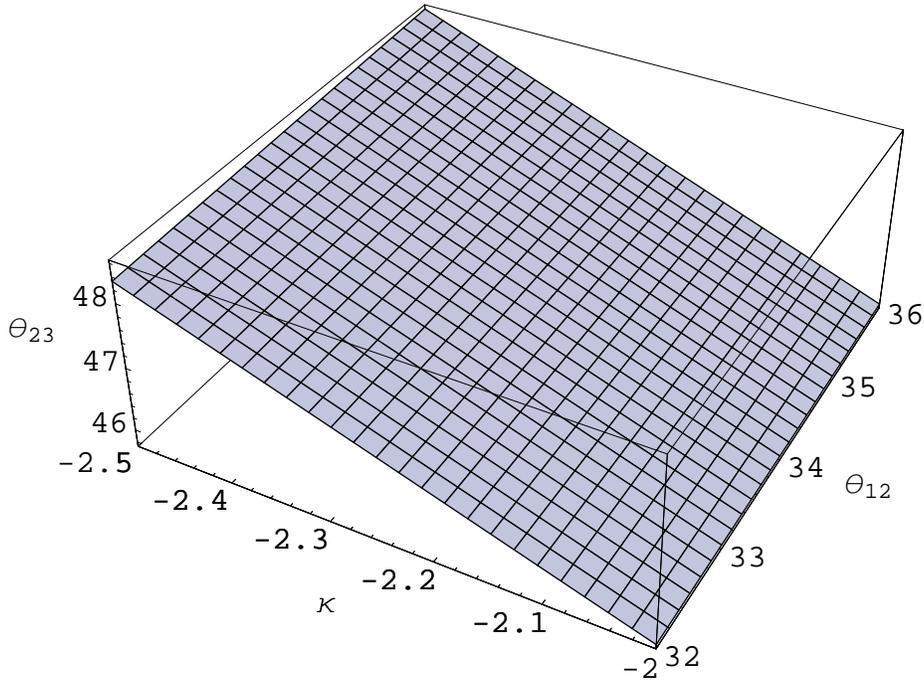}
\caption{\label{th23} Plot of   $\theta_{23}$  with respect to
  $\kappa$ and  $\theta_{12}$. We keep
  $\Delta m^2_{32}$ and  $\Delta m^2_{21}$ to their
  best fit values.  }
\end{center}
\end{figure}
%
Using the expression of $\e'$ from Eq.\ (\ref{eprime}) into the
expression of $\sin\theta_{13}$ and $\tan^2\theta_{23}$ in Eq.\ 
(\ref{angles}) we get the two mixing angles in terms of $\kappa$,
$\cos^2\phi$:
\begin{eqnarray}
 \tan^2\theta_{23} =1+\frac{2\sqrt{3}\kappa^2\cos^2\phi(\kappa+2)}{3(\kappa-2)(2+\kappa\cos^2\phi)}\times(\sin\theta_{12}-1/\sqrt{3})
\label{t23}
\end{eqnarray}
and
\begin{eqnarray}
\sin\theta_{13}=\left|\frac{\kappa\cos^2\phi(\kappa+2)}{\sqrt{6}(\kappa-2)(2+\kappa\cos^2\phi)}\right|\left[
\frac{1+\cos^2\phi(\kappa^2+8 - 6\kappa)}{\cos^2\phi}\right]^{1/2}\left|\sin\theta_{12}-1/\sqrt{3}\right|
\label{t13}
\end{eqnarray}
Here also $\tan^2\theta_{23}$ and $\sin\theta_{13}$ only depend on
$\kappa$ as $\cos^2\phi$ is only function of $\kappa$. In addition to
this dependence on parameter $\kappa$, they also depend on the well
measured quantities, mixing angle $\theta_{12}$, and the ratio of
solar and atmospheric mass-squared differences $R$. We keep the solar
and atmospheric mass-squared differences on their best fit values. We
study the variation of $\theta_{23}$ with mixing angle $\theta_{12}$
and the parameter $\kappa$ in Fig.\ref{th23}.  We have varied $\kappa$
from its analytical upper bound $-2.0$ to $-2.5$ and also have varied
$\theta_{12}$ from $32^\circ$ to $36^\circ$ ($2^\circ$ deviation about
best fit value $34^\circ$ of $\theta_{12}$). For a fixed value of
$\kappa$ variation of $\theta_{23}$ with $\theta_{12}$ is small and is
smaller in higher value of $\kappa$, e.g. for the variation of
$\theta_{12}$ from $32^\circ$ to $36^\circ$ for $\kappa=-2.0$,
$\theta_{23}$ remains almost at $45.8^\circ$ and for $\kappa=-2.5$,
$\theta_{23}$ changes from $48.4^\circ$ to $48.6^\circ$. If we allow
$\theta_{23}$ upto its $1^\circ$ deviated value $46^\circ$, range of
$\kappa$ shrinks to $-2.04\le\kappa\le -2.0$ which is insensitive to
the variation of $\theta_{12}$. To keep $\theta_{23}$ within $1\sigma$
deviated value $48^\circ$ for the whole range of variation of
$\theta_{12}$ from $32^\circ$ to $36^\circ$, we have to keep
$\kappa\ge -2.39$. From Fig.\ref{th23} we can also study the variation
of $\theta_{23}$ with $\kappa$ for a fixed value of $\theta_{12}$,
e.g. for the best fit value of $\theta_{12}=34^\circ$, $\theta_{23}$
changes from $45.8^\circ$ to $48.5^\circ$ for the variation of
$\kappa$ from $-2.0$ to $-2.5$. So, from Fig.\ref{th23} we have gathered
the information where we should keep the value of $\kappa$ in the
light of experimental values of $\theta_{23}$ and $\theta_{12}$. With
this information we  study other observables.

%
\begin{figure}
\begin{center}
\includegraphics[height=10cm,keepaspectratio]{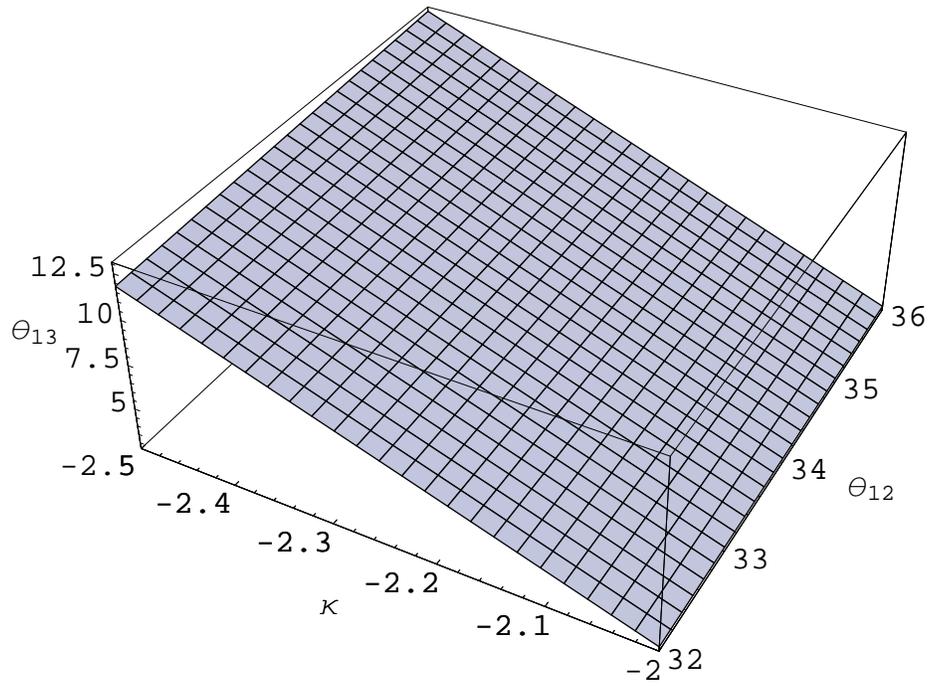}
\caption{\label{th13} Plot of   $\theta_{13}$ with respect to $\kappa$
  and $\theta_{12}$.
We keep $\Delta m^2_{32}$ and $\Delta m^2_{21}$ to their
  best fit values.  }
\end{center}
\end{figure}
Fig.\ref{th13}, we have plotted $\theta_{13}$ with $\theta_{12}$ and
$\kappa$ in their allowed region using the best fit value of solar and
atmospheric mass squared differences.  We have promised to generate nonzero
$\theta_{13}$ with changing $\theta_{12}$ from its tri-bimaximal value
$35.26^\circ$. Thus, we obtain the prediction on $\theta_{13}$ as
$2.91^\circ\le\theta_{13}\le 3.7^\circ$ for $-2.04\le\kappa\le -2.0$
(alternatively for $45.8^\circ\le\theta_{23}\le 46.0^\circ$) for the
whole range of variation of $\theta_{12}$ $32^\circ$ to $36^\circ$.
If we allow $\theta_{23}$ upto its $1\sigma$ deviated value
$45.8^\circ\le\theta_{23}\le 48.0^\circ$ (equivalently
$-2.39\le\kappa\le -2.0$), we have the prediction of $\theta_{13}$ as
$2.91^\circ\le\theta_{13}\le 10.7^\circ$ for the whole range of
variation of $\theta_{12}$  from $32^\circ$ to $36^\circ$. So, the upper
bound is near the largest possible allowed value from the experiment
(CHOOZ $\theta_{13}<12^\circ$ at $3\sigma$). From the plot
of Fig.\ref{th13}, this is to be noted that the upper bound of
$\theta_{13}$ is mildly sensitive to $\theta_{12}$, it varies from
$9.93^\circ$ to $10.7^\circ$ at $\kappa=-2.39$. But the lower bound of
$\theta_{13}$ is $2.91^\circ$ which is insensitive to the variation of
$\theta_{12}$ at $\kappa=-2.0$. 

Keeping all those constraints in view next we explore the parameter
space of CP violation parameter $J_{\rm CP}$. The parameter
$J_{\rm CP}$ defined as \cite{branco1}
\begin{eqnarray}
J_{CP} = 
  \frac{1}{8}\sin2\theta_{12}\sin2\theta_{23}\sin2\theta_{13}
  \cos\theta_{13}\sin\delta_D
   = \frac{Im[h_{12}h_{23}h_{31}]}
   {\Delta m^2_{21}\Delta m^2_{31}\Delta m^2_{32}}
\label{jcp1}
\end{eqnarray}
where $h= M_\nu M_\nu^\dagger$, $\delta_D$ is Dirac phase. This $J_{\rm
  CP}$ is associated with CP violation in neutrino oscillation and is
directly related to Dirac phase of mixing matrix. Using Eq.\ 
(\ref{mnumat}), Eq.\ (\ref{darel})  in Eq.\ 
(\ref{jcp1}) $CP$ violating parameter $J_{\rm CP}$ takes the following
form
\begin{eqnarray}
J_{\rm CP}=\frac{a^6}{\Delta m^2_{21}\Delta m^2_{32}
\Delta m^2_{31}}\times 
\frac{2\e'}{9}\kappa^3(\kappa+2)\cos^4\phi\sqrt{\cos^2\phi-\cos^4\phi}.
\label{jcp2}
\end{eqnarray}
upto first order term in $\e'$. Using the expression of $\e'$ from
Eq.\ (\ref{eprime}) and $a^2$ from Eq.\ (\ref{a}) into the Eq.\
(\ref{jcp2}) we have 
\begin{eqnarray}
J_{\rm CP} = 
\frac{2\sqrt{3}}{9}\frac{\kappa(\kappa+2)^2(2+\kappa\cos^2\phi)^2(\sin\theta_{12}-1/\sqrt{3})\sqrt{\cos^2\phi-\cos^4\phi}}{R[(\kappa-2)(2+\kappa\cos^2\phi)+2\sqrt{3}(\kappa+2)(1-\kappa\cos^2\phi)(\sin\theta_{12}-1/\sqrt{3})]^3}.
\label{jcp3}
\end{eqnarray}
%

\begin{figure}
\begin{center}
\includegraphics[height=14cm,keepaspectratio]{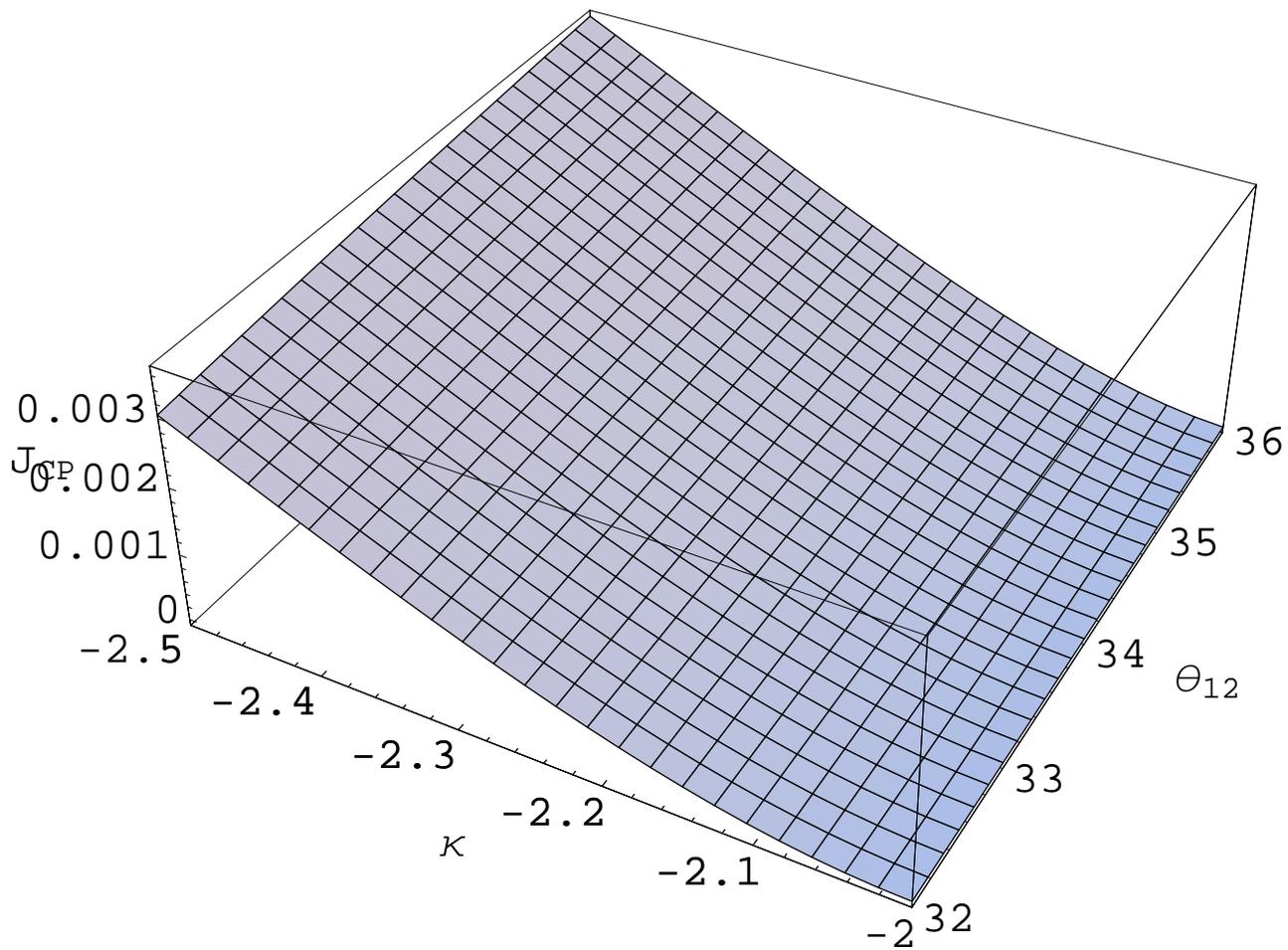}
\caption{\label{fjcp} Plot of $J_{\rm CP}$ with respect to $\kappa$
  and $\theta_{12}$. We keep
  $\Delta m^2_{32}$ and  $\Delta m^2_{21}$ to their
  best fit values.  }
\end{center}
\end{figure}
%
%

\begin{figure}
\begin{center}
\includegraphics[height=15cm,keepaspectratio]{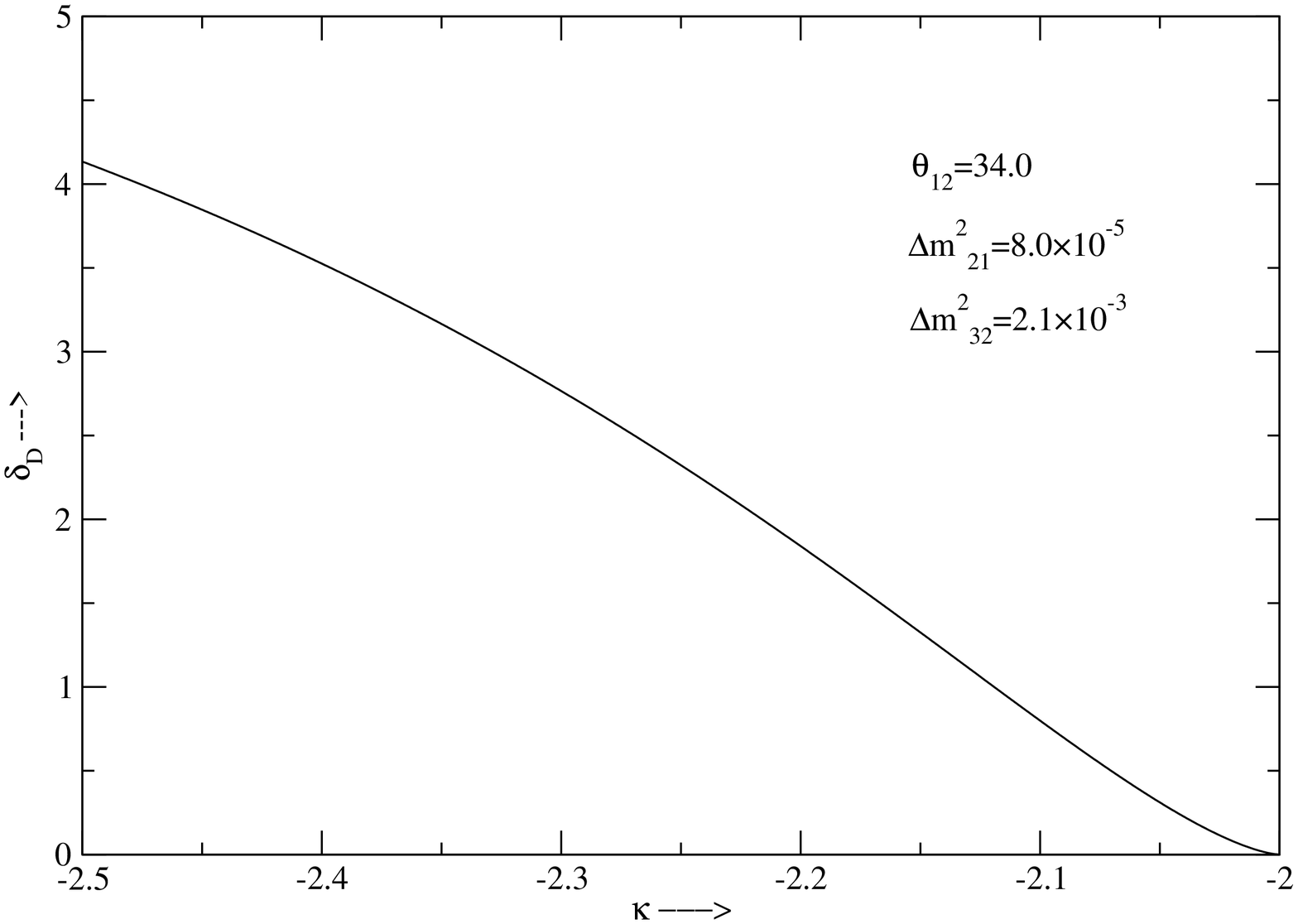}
\caption{\label{fd} Plot of Dirac phase $\delta_D$ with respect to
  $\kappa$  for the best fit values of $\theta_{12}$, $\Delta m^2_{32}$ and $\Delta m^2_{21}$. }
\end{center}
\end{figure}

$J_{\rm CP}$ is also only function of $\kappa$ because $\cos^2\phi$ in
Eq.\ (\ref{cosphi}) is function of $\kappa$ only.  It is clear from
the above expression if $\kappa = -2$ the value of $J_{\rm CP}$
becomes zero. This $\kappa+2$ factor appear in $J_{\rm CP}$ in Eq.\ 
(\ref{jcp2}) is purely from $A_4$ symmetric structure of the neutrino
mass matrix. In our analysis we have seen that $\kappa$ and
$\cos^2\phi$ are not independent.  $\kappa=-2$ approximately
corresponds to $\cos^2\phi=1$ where $J_{\rm CP}$ also vanishes. Using
the expression of $\cos^2\phi$ from Eq.\ (\ref{cosphi}) into above
Eq.\ (\ref{jcp3}) we can make $J_{\rm CP}$ only $\kappa$ dependent. It
also depends on the observables quantities $\theta_{12}$, $\Delta
m^2_{32}$ and $\Delta m^2_{21} $. We have plotted in Fig.\ref{fjcp}
$J_{\rm CP}$ with respect $\kappa$ and $\theta_{12}$ in their allowed
region keeping $\Delta m^2_{32}$ and $\Delta m^2_{21}$ to their best
fit values. We have the values of $J_{\rm CP}$ in the range
$3.27\times 10^{-5}\le J_{\rm CP}\le 7.6\times 10^{-5}$ for
$-2.04\le\kappa\le -2.0$ (alternatively for
$45.8^\circ\le\theta_{23}\le 46.0^\circ$) for the whole range of
variation of $\theta_{12}$ from $32^\circ$ to $36^\circ$. If we allow
$\theta_{23}$ upto its $1\sigma$ deviated value
$45.8^\circ\le\theta_{23}\le 48.0^\circ$ (equivalently
$-2.39\le\kappa\le -2.0$), we have the prediction of $J_{\rm CP}$ as
$3.27\times 10^{-5}\le J_{\rm CP}\le 2.65\times 10^{-3}$ for the whole
range of variation of $\theta_{12}$ from $32^\circ$ to $36^\circ$. Upper
bound of $J_{\rm CP}$ is varying from $2.12\times 10^{-3}$ to
$2.65\times 10^{-3}$ for the variation of $\theta_{12}$ from
$32^\circ$ to $36^\circ$. For the best fit value of $\theta_{12}$ ($\approx 34^\circ$), upper bound is $2.42\times 10^{-3}$. So, larger value of
$J_{\rm CP}\approx 2.65\times 10^{-3}$ is possible and it can be probed
through upcoming base-line experiments.  From the Eq.\ (\ref{jcp1}) we
can find the expression for $\sin\delta_D$. Using expressions for
$\theta_{13}$ from Eq.\ (\ref{t13}), $\theta_{23}$ from Eq.\ 
(\ref{t23}), $J_{\rm CP}$ from Eq.\ (\ref{jcp3}) and $\cos^2\phi$ from
Eq.\ (\ref{cosphi}) into Eq.\ (\ref{jcp1}) we can have $\kappa$
dependent function for $\sin\delta_D$. We have plotted $\delta_D$ with
respect to $\kappa$ in Fig.\ref{fd} for the best fit values of
$\theta_{12}$, $\Delta m^2_{32}$ and $\Delta m^2_{21}$. The figure
reflects the prediction of $\delta_D$ as $\delta_D=3.6^\circ$ for
$\theta_{23}=48^\circ$ at $\kappa=-2.39$.

\begin{figure}
\begin{center}
\includegraphics[height=15cm,keepaspectratio]{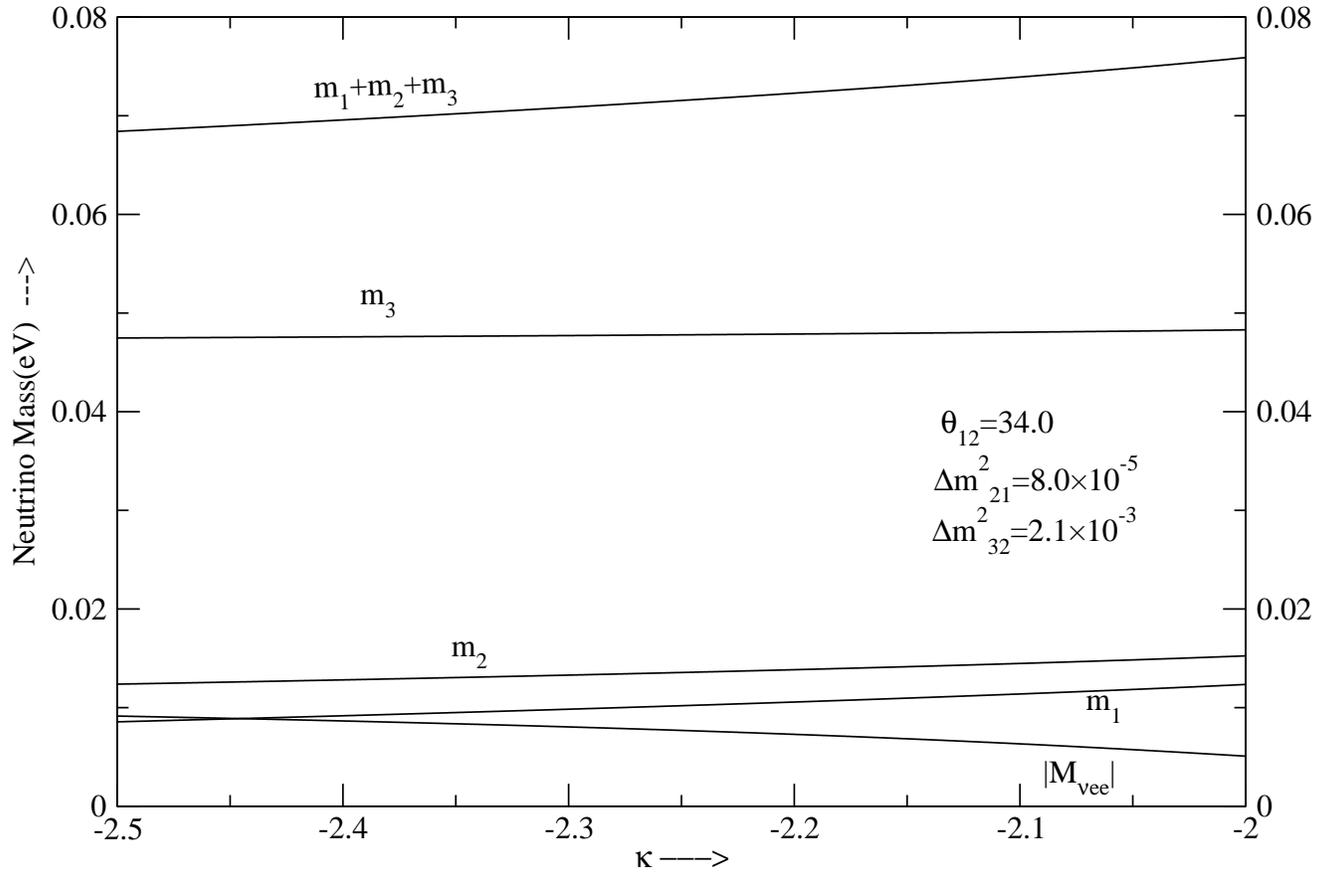}
\caption{\label{fm}  Plot of neutrino masses $m_1$, $m_2$, $m_3$ 
and their sum, and also $|(M_\nu)_{ee}|$ with respect to 
$\kappa$  for the best fit values of $\theta_{12}$, $\Delta m^2_{32}$ and $\Delta m^2_{21}$.}
\end{center}
\end{figure}

Now we are going to see the behavior of mass eigenvalues and their
sum with respect $\kappa$ and hence the other observables. Using the
the expressions for $a^2$ from Eq.\ (\ref{a}), $\e'$ from Eq.\ 
(\ref{eprime}) and $\cos^2\phi$ from Eq.\ (\ref{cosphi}) into Eq.\ 
(\ref{mev}) we get the $\kappa$ dependent functions for mass
eigen values. We have plotted $m_1$, $m_2$, $m_3$ and their sum in
Fig.\ref{fm} with respect to $\kappa$ for the best
fit values of $\theta_{12}$, $\Delta m^2_{32}$ and $\Delta m^2_{21}$.
The observations of the plots in Fig.\ref{fm} suggest that mass
pattern is normal-hierarchical. We also have seen that
$0.07~ eV<m_1+m_2+m_3< 0.076~ eV$ in the course of variation of $\kappa$ 
$-2.39<\kappa<-2$ ($48^\circ>\theta_{23}>45.8^\circ$). It also satisfy the
cosmological bound $m_1+m_2+m_3<0.7$ eV \cite{Spergel:2003cb}.

Again parameter responsible for the $\beta\beta_{0\nu}$ experiment is
also studied in the present model and the relevant quantity:
\begin{eqnarray}
\left|(M_\nu)_{ee}\right|&=&\left|a+\frac{2d\exp^{i\phi}}{3}\right|\nonumber\\
&=&a\left[1+\frac{4\kappa(\kappa+3) \cos^2\phi}{9}\right]^{1/2}.
\label{mee}
\end{eqnarray}
Using the expressions for $a$ and $\cos^2\phi$ from Eq.\ (\ref{a}) and
Eq.\ (\ref{cosphi}) respectively, we get
$\left|(M_\nu)_{ee}\right|$ in terms of model parameter $\kappa$ and
other physical observables. Keeping experimental values of the
$\theta_{12}$, $\Delta m^2_{32}$ and $\Delta m^2_{21}$ to their best
fit value, we have plotted $\left|(M_\nu)_{ee}\right|$ with respect to
$\kappa$. For the physical region of $\kappa$ Fig.\ref{fm} shows that
$\left|(M_\nu)_{ee}\right|$ is well below the experimental bound
$0.89$ eV.

A point to be noted as our analysis of mixing angles are 
matching with \cite{ba} in the real
limit($\phi=0^\circ$). However, the expressions for the mixing angles for
complex case in \cite{ba} are not in exact correspondence 
with our result in the present work for 
$\kappa=-2$ limit. This is because in the complex analysis in \cite{ba},
we have assumed that eigenvectors of $M_\nu$ construct $U$. This $U$
has ability to diagonalize $M_\nu$ to its true eigenvalues but this
$U$ may not be unitary. It is better to diagonalize $h$ ($M_\nu
M_\nu^\dagger$) which is hermitian and its diagonalizing matrix is
unitary. But it is always not easy to do the same. In this paper we
have solved the  18 equation from 
Eq.\ (\ref{md}) and find out $U$ in Eq.\
(\ref{U}) which also is unitary keeping terms upto the order
$\e$: $U^\dagger U=UU^\dagger=1+O(\e^2)$.

In summary, we explore the parameter space of a softly broken $A_4$
symmetric model for different mixing angles and a model parameter
$\kappa$. We expressed the two mixing angles $\theta_{12}$ and
$\theta_{23}$ in terms of a single parameter $\kappa$, and constrained
the parameter space for the best fit values of $\Delta m^2_{32}$ and
$\Delta m^2_{21}$. With the allowed parameter value we 
predict $\theta_{13}$, $\theta_{13}\simeq 11^\circ$ 
(for $1\sigma$ deviation of $\theta_{23}$ and 
$2^\circ$ deviation of $\theta_{12}$ about 
their best fit value). Utilising the above result, we
expressed the CP violation parameter $J_{\rm CP}$ in terms of $\kappa$
and $\sin\theta_{12}$ and explore the extent of $J_{\rm CP}$ allowed
in the present model. A comparatively larger value of $J_{\rm CP}$ is
allowed by the present model ($J_{\rm CP}$ = $2.65\times {10}^{-3}$ for
$1\sigma$ variation of the angle $\theta_{23}$) and consistent with
other neutrino experimental results.


\end{document}